\documentclass[12pt,pra,aps,amssymb,amsmath,tightenlines]{revtex4}

\usepackage{dsfont}
\usepackage[dvips]{graphicx}
\newtheorem{lem}{Lemma}
\newtheorem{prop}{Proposition}
\newcommand{\half}{\mbox{$\textstyle \frac{1}{2}$}}

\newcommand{\re}{\mbox{$\rm e$}}
\newcommand{\ri}{\mbox{$\rm i$}}
\newcommand{\rd}{\mbox{$\rm d$}}

\begin{document}
\title{Dam Rain and Cumulative Gain}
\author{Dorje~C.~Brody${}^*$,
Lane~P.~Hughston${}^{\dagger\ddagger}$, and
Andrea~Macrina${}^\dagger$}

\affiliation{${}^*$Department of Mathematics, Imperial College
London, London SW7 2BZ, UK \\ ${}^{\dagger}$Department of
Mathematics, King's College London, London WC2R 2LS, UK  \\
${}^\ddagger$Perimeter Institute, Waterloo, Ontario N2L 2YS, Canada}
\begin{abstract}
We consider a financial contract that delivers a single cash flow
given by the terminal value of a cumulative gains process. The
problem of modelling and pricing such an asset and associated
derivatives is important, for example, in the determination of
optimal insurance claims reserve policies, and in the pricing of
reinsurance contracts. In the insurance setting, the aggregate
claims play the role of the cumulative gains, and the terminal cash
flow represents the totality of the claims payable for the given
accounting period. A similar example arises when we consider the
accumulation of losses in a credit portfolio, and value a contract
that pays an amount equal to the totality of the losses over a given
time interval. An expression for the value process of such an asset
is derived as follows. We fix a probability space together with a
pricing measure, and model the terminal cash flow by a random
variable; next, we model the cumulative gains process by the product
of the terminal cash flow and an independent gamma bridge process;
finally, we take the filtration to be that generated by the
cumulative gains process. An explicit expression for the value
process is obtained by taking the discounted expectation of the
future cash flow, conditional on the relevant market information.
The price of an Arrow-Debreu security on the cumulative gains
process is determined, and is used to obtain a closed-form
expression for the price of a European-style option on the value of
the asset at the given intermediate time. The results obtained make
use of various remarkable properties of the gamma bridge process,
and are applicable to a wide variety of financial products based on
cumulative gains processes such as aggregate claims, credit
portfolio losses,  defined-benefit pension schemes, emissions, and
rainfall.
\noindent \textbf{Working paper}. This version: 15 October 2007.
\end{abstract}

\maketitle

\section{Introduction}

There are a number of problems in finance and insurance that involve
the analysis of accumulation processes---that is to say, processes
representing cumulative gains or losses. The typical setup is as
follows. We fix an accounting period $[0, T]$, where time $0$
denotes the present. At time $T$ a contract pays a random cash flow
$X_T$, which is assumed to be positive and given by the terminal
value of a process of accumulation. In the case of an insurance
contract, for example, we consider the situation where a number of
claims are made over the accounting period, and are then paid at
$T$. The random variable $X_T$ represents the totality of the
payments made at $T$ in settlement of claims arising over the
accounting period. The problem facing the insurance firm is the
valuation of the random cash flow. Let us write $\{S_t\}$  for the
value process of the contract that pays $X_T$ at $T$, and
$\{{\mathcal F}_t\}$ for the filtration representing the flow of
information available to market participants, and ${\mathbb Q}$ for
the pricing measure, which we assume to have been established by the
market. Then the value at $t$ of the contract that pays $X_T$ at $T$
is
\begin{eqnarray}
S_t = P_{tT} {\mathbb E}[X_T |{\mathcal F}_t],
\end{eqnarray}
where ${\mathbb E}[-]$ denotes expectation with respect to ${\mathbb
Q}$, and $P_{tT}$ denotes the discount factor, which for simplicity
we take to be deterministic. One can interpret $S_t$ as the reserve
that the insurance firm requires at $t$ to ensure that $X_T$ will be
payable at $T$. Alternatively, one can view $S_t$ as the amount that
would have to be paid at $t$ in order for the insurance firm to
relieve itself of the obligation to pay $X_T$, that is to say, to
commute the relevant claims.  Similarly, the cost $C_{tT}$ at $t$ of
a simple stop-loss reinsurance contract that pays out $(X_T - K)^+$
at $T$ for some fixed threshold $K$ is given by
\begin{eqnarray}
C_{tT} = P_{tT} {\mathbb E}[(X_T - K)^+ |{\mathcal F}_t].
\end{eqnarray}

We shall assume that $\{{\mathcal F}_t\}$ is generated by an
aggregate claims process $\{\xi_t\}$, where for each $t$ the random
variable $\xi_t$ represents the totality of claims known at $t$ to
be payable at $T$. The problem can then be stated as follows: given
the history of claims over the accounting period up to time $t$,
what is the appropriate reserve to allocate for settlement of these
and any future claims arising in the accounting period? To obtain a
solution to the problem we need to specify the aggregate
claims process, then work
out the reserve process $\{S_t\}$. Once we have the reserve process,
we can value various types of reinsurance contracts.

Another example of an accumulation process comes from credit risk
management. We consider a large credit portfolio, and let $X_T$
denote the value of the accumulated losses at $T$. For instance, at
time $0$ a credit-card firm has a large number of customers, each
with an outstanding balance payable in the accounting period. If a
customer does not pay the balance by the required date, they will be
deemed to be in default, and a loss will be registered. The random
variable $X_T$ will denote the totality of such losses. We assume
that once a customer is in default, no further payments are made by
that customer (this assumption can be relaxed in a more
sophisticated model). The problem facing the credit-card firm is to
determine what reserve policy to maintain, and what premium to
charge over the base interest rate, to ensure that funds will be in
hand to cover the default losses.

The purpose of this paper is to present a modelling framework for
accumulation processes, and to establish explicit formulae for the
associated valuation processes. In particular, we shall assume that
$\{\xi_t\}$ takes the form
\begin{eqnarray}
\xi_t = X_T \gamma_{tT}, \label{eq:3.0}
\end{eqnarray}
where $\{\gamma_{tT}\}$ is a gamma bridge over the interval $[0,T]$,
independent of $X_T$. The motivation for the form of the
accumulation process indicated above arises in two distinct lines of
enquiry. The first relates to the idea that the gamma process might
be used as a basis for describing the aggregate losses associated
with insurance claims. This idea dates to the work of Hammersley
(1955), Moran (1956), Gani (1957), Kendall (1957), and others, in
connection with the theory of storage and dams. Moran (1956), in
particular, observed that the amount of rainfall accumulating in a
dam can be modelled by a gamma process, and Gani (1957) pointed out
the relevance to insurance, the argument being that providing that
the portfolio of events insured is sufficiently large, one can think
of the arrival of claims as being analogous to the accumulation of
dam rain. The gamma process has since then been investigated by
Dufresne \textit{et al}. (1991), Dufresne (1998), Dicksen \& Waters
(1993), and others, as a model for aggregate claims.

Let us therefore consider what results if we  model the aggregate
claims process as a ${\mathbb Q}$-gamma process. In other words,
suppose we set $\xi_t=\kappa \gamma_t$, where $\kappa $ is a
constant and  $\{\gamma_t\}$ is a standard gamma process under
${\mathbb Q}$, with mean and variance $mt$ (see Section~\ref{sec:2}
for definitions). It follows that $\xi_t = X_T \gamma_{tT}$ where
$X_T = \kappa \gamma_T$ and the process $\{\gamma_{tT}\}$ defined by
$\gamma_{tT} = \gamma_t / \gamma_T$ is a standard gamma bridge over
$[0,T]$. Moreover, by virtue of the special properties of the gamma
process, we find that $X_T$ is independent of $\{\gamma_{tT}\}$. We
see that in the ${\mathbb Q}$-gamma model the aggregate claims
process is the product of a gamma-distributed terminal cash flow and
an independent gamma bridge. One can think of the gamma bridge as
representing that aspect of the aggregate claims process that has no
bearing on the terminal result. We are thus led to a multiplicative
decomposition of the accumulation process into the product of a
``signal" $X_T$ and an independent ``noise" $\{\gamma_{tT}\}$
carrying no information about $X_T$.

For such processes we are able to apply the techniques of
information-based asset pricing developed in Brody \textit{et al}.
2007a, 2007b, Hughston \& Macrina 2007, Macrina 2006; and Rutkowski
\& Yu 2007. Indeed, through this second line of enquiry one is led
to consider the more general situation where the terminal cash flow,
instead of being gamma distributed, has a generic \textit{a priori}
distribution, and the claims process takes the form (\ref{eq:3.0}).
The additive decomposition of the market information process in the
case of the Brownian bridge noise considered in the references cited
above is natural from the viewpoint of nonlinear filtering theory.
The product representation of the gamma information process is
equally natural, since many properties of the Brownian bridge that
hold additively have striking multiplicative analogues for gamma
bridges (Emery \& Yor 2004, Yor 2007). The resulting model for the
aggregate claims process is remarkably tractable, and we are able to
derive explicit formulae both for the claims reserve process, and
for the valuation of reinsurance contracts.

The paper is organised as follows. In Sections II, III, and IV, we
outline a number of the properties of gamma processes and gamma
bridges. The material covered in these sections is for the most part
well known. However, since it is not easy to locate a systematic but
elementary treatment of the gamma process and the associated bridge
process, it will be useful to present some of the details here for
the benefit of general readers. At the same time, we establish our
notation and some results that will be applied in later sections. In
Section~\ref{sec:5} we derive an explicit expression for the value
process of a contract that delivers the cash flow $X_T$ at time $T$,
when the market filtration is generated by the accumulation process
(\ref{eq:3.0}). We show in Proposition~\ref{prop:4} that $\{\xi_t\}$
has the Markov property, and then use the Bayes theorem to determine
the conditional density of $X_T$, and finally the value process,
which is given in Proposition~\ref{prop:5}. By use of the
conditional density we are also able to obtain an expression for the
value process of a simple stop-loss reinsurance contract. In
Section~\ref{sec:6} we consider the valuation of general reinsurance
contracts. In particular, we derive a formula for the value at time
$0$ of a contract that at some fixed time $t$ gives the contract
holder the option to commute the claim $X_T$ by paying a fixed
amount $K$ at $t$. Such a contract takes the form of a European call
option on the value of the reserve at $t$. An Arrow-Debrue method is
introduced to simplify the calculations. The resulting formula for
the option value is expressed in terms of the cumulative beta
distribution. We examine in Section~\ref{sec:7} the case where $X_T$
takes discrete values. When $X_T$ is a binary random variable, the
problem of option pricing can be solved completely. In
Section~\ref{sec:8} the material of Section~\ref{sec:6} is extended
to determine an expression for the price process of an option on the
value of an aggregate claim. In Section~\ref{sec:9} we conclude by
returning to the case where $X_T$ has a ${\mathbb Q}$-gamma
distribution.

\section{Gamma processes and associated martingales}
\label{sec:2}

We fix a probability space $(\Omega,{\mathcal F},{\mathbb Q})$. In
our applications ${\mathbb Q}$ will always denote the pricing
(risk-neutral) measure, but the material in this section, and the
following two, does not depend on this interpretation. Equalities
and inequalities among random variables are to be understood as
holding except possibly on sets of measure zero. By a \emph{standard
gamma process} $\{\gamma_t\}_{0\le t< \infty}$ on $(\Omega,{\mathcal
F},{\mathbb Q})$ with growth rate $m$ we mean a process with
independent increments such that $\gamma_0=0$ and such that the
random variable $\gamma_t$ has a gamma distribution with mean and
variance $mt$. More precisely, writing $G(x)={\mathbb Q}[\gamma_t\le
x]$ for the distribution of $\gamma_t$, and writing $g(x)=\rd
G(x)/\rd x$, we have
\begin{eqnarray}
g(x)={\mathds 1}_{\{x>0\}} \frac{x^{mt-1}\re^{-x}}
{\Gamma[mt]} \label{eq:gamma}
\end{eqnarray}
for the density of $\gamma_t$. Here $\Gamma[a]$
is the standard gamma function, which for $a>0$ has the
Eulerian representation:
\begin{eqnarray}\label{gammafct}
\Gamma[a]=\int^{\infty}_0\,x^{a-1}\re^{-x}\rd x.
\end{eqnarray}
It follows from the identity $\Gamma[a+1]=a\Gamma[a]$
satisfied by the gamma function that
\begin{eqnarray}
{\mathbb E}[\gamma_t]=mt,
\end{eqnarray}
which justifies the interpretation of the parameter $m$ as the
\emph{mean growth rate} of the process.

A straightforward calculation shows that the characteristic function
for the gamma process is given by
\begin{eqnarray}
{\mathbb E}\left[\re^{{\rm i}\lambda\gamma_t}\right]=
\frac{1}{(1-{\rm i}\lambda)^{mt}}, \label{eq:1.4}
\end{eqnarray}
valid for $t\ge 0$ and for $\lambda\in\mathbb{C}$ such that ${\rm
Im} (\lambda)>-1$, from which the higher moments of $\gamma_t$ can
be deduced. We note that ${\mathbb E} \left[ \gamma_t^2
\right]=mt+m^2 t^2$, and hence that $\textrm{Var}[\gamma_t]=mt$. It
follows as a consequence of the independent increments property that
$\textrm{Cov}[\gamma_t,\gamma_u]=mt$ for $u\geq t$.

An alternative expression for the characteristic function is given
by the L\'evy-Khinchine representation ${\mathbb E}\left[\re^{{\rm
i}\lambda\gamma_t} \right] = \re^{-t \psi(\lambda)}$ for ${\rm Im}
(\lambda)>-1$, where
\begin{eqnarray}
\psi(\lambda) = m \ln(1 - \ri \lambda) = \int_0^\infty
mx^{-1}\re^{-x} \left(1-\re^{{\rm i} \lambda x} \right) \rd x,
\end{eqnarray}
which shows that the L\'evy density associated with the gamma
process is given by $m x^{-1} \re^{-x}$ for $x>0$ (see, e.g.,
Protter 2005).

By use of the independent increments property we deduce that for
$u\ge t\ge 0$ and for $a,b\in\mathbb{C}$ with ${\rm Im}(a+b)>-1$
and ${\rm Im}(b)>-1$ we have:
\begin{eqnarray}
{\mathbb E}\left[\re^{{\rm i} a\gamma_t+{\rm i} b \gamma_u} \right]
&=&{\mathbb E}\left[\re^{{\rm i}(a+b)\gamma_t+{\rm i}
b(\gamma_u-\gamma_t)}\right]\nonumber\\ &=&{\mathbb E}
\left[\re^{{\rm i}(a+b)\gamma_t}\right]{\mathbb E} \left[\re^{{\rm
i} b(\gamma_u-\gamma_t)}\right] \nonumber\\
&=&\frac{1}{[1-\ri(a+b)]^{mt}}\ \frac{1}{(1-\ri\,b)^{m(u-t)}}.
\label{eq:6}
\end{eqnarray}
In particular if we set $-a = b = \lambda$, we see that $\gamma_u-
\gamma_t$ is gamma-distributed with parameter $m(u-t)$. It follows
that the increments of $\{\gamma_t\}$ have a time-homogeneous
probability law in the sense that  $\gamma_{u+h}- \gamma_{t+h}$ has
the same distribution as $\gamma_u- \gamma_t$.

Using the independent increments property it is a straightforward
exercise to deduce that the processes $\{\gamma_t-mt\}$ and
$\{\gamma_t^2-2mt\gamma_t+mt(mt-1)\}$ are martingales. More
generally, for $\alpha>-1$ the process $\{L_t\}$ defined by
\begin{eqnarray}
L_t = (1+\alpha)^{mt}\re^{-\alpha\gamma_t} \label{eq:ww1}
\end{eqnarray}
is a martingale, which can be verified by use of (\ref{eq:6}). We
refer to this process as the exponential gamma martingale. It
follows, by consideration of the corresponding power series in
$\alpha$, that for each term in the series we are able to obtain a
martingale involving a polynomial expression in the gamma process.
Suppose for $n\in{\mathds N}$ and $k\in{\mathds R}$ we define the
so-called associated Laguerre polynomials $\{L_n^{(k)}(z)\}$ by
setting
\begin{eqnarray}
L_n^{(k)}(z) = z^{-k}\re^{z} \frac{\rd^n}{\rd z^n}\, \big( z^{n+k}
\re^{-z} \big).
\end{eqnarray}
Thus, we have $L_1^{(k)}(z) = -z+k+1$, $L_2^{(k)}(z) = \half
[z^2-2(k+2)z + (k+1)(k+2)]$, and so on. The standard Laguerre
polynomials, given by $L_n(z)=L_n^{(0)}(z)$, have the property that
if $Z$ is a standard exponentially distributed random variable, then
${\mathbb E}\left[ L_n(Z)L_{n'}(Z)\right] = 0$ for $n\neq n'$ (cf.
Wiener 1949). More generally, if $Z$ has a gamma distribution with
parameter $k+1$, i.e. such that ${\mathbb Q}[Z<z]=\int_0^z x^k
\re^{-x} \rd x / \Gamma[k+1]$, for $k>-1$, then ${\mathbb E}[
L_n^{(k)}(Z) L_{n'}^{(k)}(Z)] = 0$ for $n\neq n'$. The significance
of the associated Laguerre polynomials in the present context arises
from the identity
\begin{eqnarray}
(1+\alpha)^{h} \re^{-z\alpha} = \sum_{n=0}^\infty L_n^{(h-n)}(z) \,
\alpha^n, \label{eq:Lag}
\end{eqnarray}
valid for $|\alpha|<1$ and $h\geq 0$ (Erd\'elyi 1953), which gives
us the required series expansion of the exponential gamma martingale
in powers of $\alpha$. In particular, by setting $h=mt$ and
$z=\gamma_t$ in equation (\ref{eq:Lag}), we are able to deduce that
for each value of $n$ the process $\{L_n^{(mt-n)}(\gamma_t)\}$ is a
martingale (cf. Schoutens 2000). For example, we have
\begin{eqnarray}
L_1^{(mt-1)}(\gamma_t) = - (\gamma_t-mt),  \qquad
L_2^{(mt-2)}(\gamma_t) = \half [\gamma_t^2 -2mt\gamma_t +
mt(mt-1)].
\end{eqnarray}

So far we have confined the discussion to the case of the
``standard'' gamma process, for which ${\mathbb E}[\gamma_t]=mt$ and
${\rm Var}[\gamma_t]=mt$, for some value of $m$. We note that the ratio
$({\mathbb E}[\gamma_t])^2/{\rm Var}[\gamma_t]$ is dimensionless,
and hence that $m$ has the units of inverse time. For any fixed $m$
we can choose the units of time so that $m=1$ in those
units (this is done implicitly, for example, in Yor 2007). We shall,
however, take the units of time as fixed, and $m$
as a model parameter.

For many applications it is useful also to consider a broader
family of gamma processes, labelled by two parameters, which we
shall call ``scaled'' gamma processes. By a scaled gamma process
with growth rate $\mu$ and spread $\sigma$ we mean a process
$\{\Gamma_t\}_{0 \leq t<\infty}$ with independent increments such
that $\Gamma_0=0$ and such that $\Gamma_t$ has a gamma
distribution with mean $\mu t$ and variance $\sigma^2t$, where
$\mu$ and $\sigma$ are parameters. Defining $m=\mu^2/\sigma^2$ and
$\kappa=\sigma^2/\mu$, we have $\mu= \kappa m$ and
$\sigma^2=\kappa^2 m$. One can think of $m$ as a ``standardised''
growth rate, and $\kappa$ as a ``scale''. The density of
$\Gamma_t$ is then given by
\begin{eqnarray}
g_{\Gamma_t}(x) = {\mathds 1}_{\{x>0\}} \frac{\kappa ^{-mt}
x^{mt-1} \re^{-x/\kappa }}{\Gamma[mt]}. \label{eq:ww2}
\end{eqnarray}
It is straightforward to check that if $\{\Gamma_t\}$ is a scaled
gamma process with standardised growth rate $m$ and scale $\kappa$,
then $\{\kappa^{-1}\Gamma_t\}$ is a standard gamma process, with
growth rate $m$.

Now suppose that $\{\gamma_t\}$ is a standard gamma process on
$(\Omega,{\mathcal F},{\mathbb Q})$, let $\{{\mathcal G}_t\}$
denote the filtration generated by $\{\gamma_t\}$, and let
${\mathbb Q}^*$ denote the measure on $(\Omega,{\mathcal G}_T)$,
for some fixed $T$, defined by the likelihood ratio
\begin{eqnarray}
\left.\frac{\rd{\mathbb Q}^*}{\rd{\mathbb Q\,\,}}\right|_T=
\kappa^{-mT} \exp\left( \frac{1-\kappa}{\kappa}\,\gamma_T\right)
\end{eqnarray}
for some $\kappa>0$. Then $\{\gamma_t\}_{0\leq t\leq T}$ is a
\textit{scaled} gamma process on $(\Omega,{\mathcal G}_T,{\mathbb
Q}^*)$, with scale parameter $\kappa$. Thus, ${\mathbb E}[\gamma_t]
=mt$, ${\rm Var}[\gamma_t]=mt$, ${\mathbb E}^*[\gamma_t]=\kappa mt$,
and ${\rm Var}^*[\gamma_t]=\kappa^2mt$. This can be established by
working out the joint characteristic function under ${\mathbb Q}^*$
of the increments $\gamma_t-\gamma_s$, $\gamma_s-\gamma_{s_1}$,
$\gamma_{s_1}-\gamma_{s_2}$, $\cdots$, $\gamma_{s_{n-1}}-
\gamma_{s_n}$ for $T\geq t\geq s\geq s_1\geq s_2\geq \cdots \geq
s_n$ for each $n\in{\mathds N}$, and showing that it factorises. We
note that the change-of-measure density martingale arising in this
example is obtained by taking the standard gamma exponential
martingale (\ref{eq:ww1}) defined above, and setting $\alpha=(1-
\kappa)/\kappa$.

The gamma process has been used as the basis of a number of
different asset pricing models; see, for example, Madan \& Seneta
(1990), Madan and Milne (1991), Heston (1995), Madan \textit{et al}.
(1998), Carr \textit{et al}. (2002), and Baxter (2007).

\section{Gamma bridge processes}
\label{sec:3}

Let $\{\gamma_t\}_{0\le t<\infty}$ be a standard gamma
process with growth rate $m$, and for fixed $T$ define the process
$\{\gamma_{tT}\}_{0\le t\le T}$ by setting
\begin{eqnarray}
\gamma_{tT}=\frac{\gamma_t}{\gamma_T}.
\end{eqnarray}
Then clearly $\gamma_{0T}=0$ and $\gamma_{TT}=1$. We refer to
$\{\gamma_{tT}\}$, thus defined, as the \emph{standard gamma bridge}
over $[0,T]$ associated with the gamma process $\{\gamma_t\}$. More
generally, we refer to any process having the law of
$\{\gamma_{tT}\}$ as a standard gamma bridge over $[0,T]$. It can be
shown that the random variable $\gamma_{tT}$ has a beta
distribution. In particular, we have the following:

\begin{prop}\label{prop:1}
The density function of the random variable $\gamma_{tT}$ is given
by
\begin{eqnarray}\label{prop1}
f(y)={\mathds 1}_{\{0<y<1\}}
\frac{y^{mt-1}(1-y)^{m(T-t)-1}}{\textrm{B}[mt,m(T-t)]},
\end{eqnarray}
where
\begin{eqnarray}\label{prop1a}
\textrm{B}[a,b]=\frac{\Gamma[a]\Gamma[b]}{\Gamma[a+b]}.
\end{eqnarray}
\end{prop}

\noindent{\bf Proof}. First we note that
\begin{eqnarray}
{\mathbb Q}\left[\frac{\gamma_{t}}{\gamma_T}\le y\right]={\mathbb
Q}\left[\frac{\gamma_t}{\gamma_T-
\gamma_t}\le\frac{y}{1-y}\right].
\end{eqnarray}
Since $\gamma_t$ and $\gamma_T-\gamma_t$ are independent, and
$\gamma_t$ has a gamma distribution with parameter $mt$, we have
\begin{eqnarray}
{\mathbb Q}\left[\frac{\gamma_t}{\gamma_T-\gamma_t}\le
\frac{y}{1-y}\right]&=&{\mathbb Q}\left[\gamma_t\le
\frac{y}{1-y}(\gamma_T-\gamma_t)\right]\nonumber\\
&=&{\mathbb E}\left[{\mathbb Q}\left[\gamma_t\le\frac{y}{1-y}
(\gamma_T-\gamma_t)\bigg\vert\gamma_T-\gamma_t\right]\right]
\nonumber\\ &=&\frac{1}{\Gamma[mt]}\,{\mathbb
E}\left[\int^{\frac{y}{1-y}
(\gamma_T-\gamma_t)}_0\,x^{mt-1}\re^{-x}\rd x\right].
\end{eqnarray}
Therefore, the corresponding density is given by
\begin{eqnarray}
f(y)&=&\frac{\rd}{\rd y}{\mathbb Q}
\left[\frac{\gamma_t}{\gamma_T}\le y \right] \nonumber \\
&=&{\mathds 1}_{\{0<y<1\}} \frac{1}{\Gamma[mt]} \, {\mathbb
E}\left[\frac{\rd}{\rd y}\int^{\frac{y}{1-y}(\gamma_T-\gamma_t)}_0
x^{mt-1}\re^{-x}\,\rd x\right] \nonumber \\ &=&{\mathds
1}_{\{0<y<1\}}\frac{1}{\Gamma[mt]}\,{\mathbb E}\left[
\frac{\gamma_T-\gamma_t}{(1-y)^2}\left(\frac{y}{1-y}(
\gamma_T-\gamma_t)\right)^{mt-1}\re^{-\frac{y}{1-y}
(\gamma_T-\gamma_t)}\right] \nonumber \\ &=& {\mathds
1}_{\{0<y<1\}}\frac{y^{mt-1}(1-y)^{-mt-1}}{\Gamma[mt]}\,{\mathbb E}
\left[(\gamma_T-\gamma_t)^{mt}\re^{-\frac{y}{1-y}
(\gamma_T-\gamma_t)}\right]. \label{eq:13}
\end{eqnarray}
Now, since $\gamma_T-\gamma_t$ has a gamma
distribution with parameter $m(T-t)$, for the expectation appearing
in the line just above we obtain
\begin{eqnarray}
{\mathbb E}\left[(\gamma_T-\gamma_t)^{mt}\re^{-\frac{y}{1-y}
(\gamma_T-\gamma_t)} \right] &=& \frac{1}{\Gamma[m(T-t)]}
\int^{\infty}_0 x^{mt} \re^{-\frac{y}{1-y}x}x^{m(T-t)-1} \re^{-x}\rd
x\nonumber\\ &=& \frac{1}{\Gamma[m(T-t)]} \int^{\infty}_0 x^{mT-1}
\re^{-\frac{y}{1-y}x}\rd x\nonumber\\ &=&
\frac{(1-y)^{mT}}{\Gamma[m(T-t)]} \int^{\infty}_0
u^{mT-1}\re^{-u}\rd u\nonumber\\ &=&
\frac{\Gamma[mT]}{\Gamma[m(T-t)]}\, (1-y)^{mT},
\end{eqnarray}
where in the last two steps we make the substitution $x=u(1-y)$ and
use formula (\ref{gammafct}). Putting this result back into
(\ref{eq:13}), we obtain (\ref{prop1}), as desired.\hfill$\Box$
\vspace{0.4cm}

Let us calculate the moments of $\gamma_{tT}$. Bearing in mind the
integral representation
\begin{eqnarray}
\textrm{B}[a,b]=\int^1_0\,y^{a-1}(1-y)^{b-1}\rd y
\end{eqnarray}
for the beta function, we deduce that
\begin{eqnarray}
{\mathbb E}\left[\gamma_{tT}^{n}\right] = \frac{\textrm{B}
[mt+n,m(T-t)]}{\textrm{B}[mt,m(T-t)]} \label{eq:3.17}
\end{eqnarray}
for $n>0$. By use of (\ref{prop1a}) along with the identity
$\Gamma[a+1]=a\Gamma[a]$ we find that ${\mathbb E}[\gamma_{tT}]=t/T$
and that ${\mathbb E}[\gamma_{tT}^2]=t(mt+1)/T(mT+1)$. It follows in
particular that
\begin{eqnarray}
\textrm{Var}[\gamma_{tT}]=\frac{t(T-t)}{T^2(1+mT)}.
\end{eqnarray}
It is interesting to observe that the expectation of $\gamma_{tT}$
does not depend on the growth rate $m$, and that the variance of
$\gamma_{tT}$ decreases in increasing $m$.

More generally, let us define the Pochhammer symbol by writing
$(a)_0=1$ and $(a)_k =a(a+1)(a+2) \cdots (a+k-1)$. Then we find that
the moments of $\gamma_{tT}$ are given by the expression ${\mathbb
E} [\gamma_{tT}^{n}]= (mt)_n/(mT)_n$, and for the corresponding
central moments we obtain
\begin{eqnarray}
{\mathbb E}\left[(\gamma_{tT}-{\mathbb E}[\gamma_{tT}])^n\right] =
\left(-\frac{t}{T}\right)^n F\left( -n,mt;mT;T/t\right),
\label{eq:3.176}
\end{eqnarray}
where $F\left(a,b;c;z\right) = \sum_{k=0}^\infty (a)_k (b)_kz^k /
[k!(c)_k]$ is the hypergeometric function (Erd\'elyi 1953).

\section{Further properties of gamma bridges}
\label{sec:4}

It is a remarkable property of the gamma process and the associated
gamma bridge that the processes $\{\gamma_u\}_{u\ge T}$ and
$\{\gamma_{tT}\}_{0\le t\le T}$ are independent. In particular, the
random variables $\gamma_T$ and $\gamma_{tT}=\gamma_t/\gamma_T$ are
independent for $0\le t\le T$. This property allows us to verify
straightforwardly that $\{\gamma_t\}$ has the Markov property.
To show that $\{\gamma_t\}$ has the Markov property we
need to verify for $a>0$ that
\begin{eqnarray}
{\mathbb Q}\left[\gamma_t<a\vert\gamma_s,\gamma_{s_1},
\gamma_{s_2},\ldots,\gamma_{s_n}\right]={\mathbb Q}
[\gamma_t<a\vert\gamma_s]
\end{eqnarray}
for all $t\ge s\ge s_1\ge s_2\ge\cdots\ge s_n\ge 0$, and for all
$n\ge1$. But clearly,
\begin{eqnarray}
{\mathbb Q}\left[\gamma_t<a\vert\gamma_s,\gamma_{s_1},
\gamma_{s_2},\ldots,\gamma_{s_n}\right]&=&{\mathbb Q}
\left[\gamma_t<a\,\bigg\vert\,\gamma_s,
\frac{\gamma_{s_1}}{\gamma_s},\frac{\gamma_{s_2}}{\gamma_{s_1}},
\ldots\right]\nonumber\\
&=&{\mathbb Q}\left[\gamma_t<a\vert\gamma_s\right],
\end{eqnarray}
since, according to the result to be established below,
$\gamma_{s_1}/ \gamma_s$, $\gamma_{s_2}/\gamma_{s_1},\ldots$ are
independent of $\gamma_s$ and $\gamma_t$. It follows that the gamma
process is Markovian. A similar argument shows that the gamma bridge
has the Markov property. In particular, we have
\begin{eqnarray}
{\mathbb Q}\left[\frac{\gamma_t}{\gamma_T}<a\,\bigg\vert\,
\frac{\gamma_s}{\gamma_T},\frac{\gamma_{s_1}}{\gamma_T},
\frac{\gamma_{s_2}}{\gamma_T},\ldots\right] &=&{\mathbb
Q}\left[\frac{\gamma_t}{\gamma_T}<a\,\bigg\vert\,
\frac{\gamma_s}{\gamma_T},\frac{\gamma_{s_1}}{\gamma_s},
\frac{\gamma_{s_2}}{\gamma_{s_1}},\ldots\right]\nonumber\\
&=&{\mathbb Q}\left[\frac{\gamma_t}{\gamma_T}<a\,\bigg\vert\,
\frac{\gamma_s}{\gamma_T}\right],
\end{eqnarray}
since the random variables $\gamma_{s_1}/\gamma_s$,
$\gamma_{s_2}/\gamma_{s_1},\ldots$ are independent of
$\gamma_t/\gamma_T$ and $\gamma_s/\gamma_T$.

\begin{prop} \label{prop:2}
Let $\{\gamma_t\}_{0\le t<\infty}$ be a standard gamma process. Then
for $T\ge t\ge 0$ the random variables $\gamma_t/\gamma_T$ and
$\gamma_T$ are independent.
\end{prop}

\noindent{\bf Proof}. For the joint distribution of these
random variables let us write
\begin{eqnarray}
F(y,z)={\mathbb Q}\left[\frac{\gamma_t}{\gamma_T}\le y\ \cap\
\gamma_T\le z\right].
\end{eqnarray}
We note that this can be rearranged in the form
\begin{eqnarray}
F(y,z)={\mathbb Q} \left[\gamma_t \le\frac{y}{1-y} (\gamma_T-
\gamma_t)\ \cap\ \gamma_t\le z-(\gamma_T-\gamma_t)\right].
\end{eqnarray}
Conditioning with respect to $\gamma_T-\gamma_t$, we use the
fact that $\gamma_t$ and $\gamma_T-\gamma_t$ are independent, and
that $\gamma_t$ has a gamma distribution with parameter $mt$, to
deduce that
\begin{eqnarray}
F(y,z)=\frac{1}{\Gamma[mt]}\,{\mathbb E}\left[ \int^{\infty}_0
{\mathds 1}_{\left\{x\le\frac{y}{1-y} (\gamma_T-\gamma_t)\right\}}
{\mathds 1}_{\left\{x\le z-(\gamma_T-\gamma_t)\right\}}
x^{mt-1}\re^{-x}\rd x\right].
\end{eqnarray}
Differentiating each side of this relation with respect to $y$ and
$z$, we obtain the following expression for the joint density
function:
\begin{eqnarray}
f(y,z)&=&\frac{1}{\Gamma[mt]}\,{\mathbb E}\left[ \int_0^\infty
\frac{\gamma_T-\gamma_t}{(1-y)^2}\, \delta\! \left(
x-\frac{y}{1-y} (\gamma_T-\gamma_t)\right) \right. \nonumber
\\ && \hspace{3.0cm} \times\  \delta \big(x- [z-(\gamma_T-\gamma_t)]
\big) x^{mt-1}\re^{-x}\rd x \Big] .
\end{eqnarray}
Here we have used the relation $\partial_x {\mathds 1}_{\{x\le
a\}}=-\delta(x-a)$, where $\{\delta(z)\}_{z\in{\mathds R}}$ denotes
the Dirac distribution. Integrating out the first delta function we
thus have
\begin{eqnarray}
f(y,z)&=&\frac{1}{\Gamma[mt]}\,{\mathbb E}\left[
\frac{\gamma_T-\gamma_t}{(1-y)^2}\, \delta \! \left(
\frac{y}{1-y}(\gamma_T-\gamma_t)- [z-(\gamma_T-\gamma_t)]
\right)\right. \nonumber \\
&&\hspace{4cm}\times\left.\left(\frac{y}{1-y}
(\gamma_T-\gamma_t)\right)^{mt-1}\re^{-\frac{y}{1-y}
(\gamma_T-\gamma_t)}\right],
\end{eqnarray}
for $0< y< 1$ and $z>0$; and hence after some rearrangement we
obtain
\begin{eqnarray}
f(y,z)=\frac{y^{mt-1}(1-y)^{-mt-1}}{\Gamma[mt]}\,{\mathbb E} \left[
(\gamma_T-\gamma_t)^{mt}\re^{-\frac{y}{1-y} (\gamma_T-\gamma_t)}\
\delta\!\left( \frac{\gamma_T-\gamma_t}{1-y} -z\right)\right].
\end{eqnarray}
Now we introduce the Fourier representation
\begin{eqnarray}\label{deltarep}
\delta(x)=\frac{1}{2\pi}\int_{-\infty}^{\infty}\re^{{\rm i}\lambda
x}\rd\lambda
\end{eqnarray}
for the delta function, interpreted in a distributional sense,
from which we deduce that
\begin{eqnarray}\label{proof2-eq1}
f(y,z)=\frac{y^{mt-1}(1-y)^{-mt-1}}{\Gamma[mt]}
\frac{1}{2\pi}\int^{\infty}_{-\infty} \re^{-{\rm i}\lambda
z}{\mathbb E}\left[(\gamma_T-\gamma_t)^{mt}\re^{-\frac{y}{1-y}
(\gamma_T-\gamma_t)}\re^{{\rm i} \lambda\frac{1}{1-y}
(\gamma_T-\gamma_t)}\right]\rd\lambda.
\end{eqnarray}
Writing $E$ for the result of the expectation appearing inside the
integral above, and making use of the fact that
$\gamma_T-\gamma_t$ is gamma distributed with parameter $m(T-t)$,
we have
\begin{eqnarray}
E&=&\frac{1}{\Gamma[m(T-t)]}\int^{\infty}_0\,x^{mt}
\re^{-\frac{y}{1-y}x}\re^{{\rm
i}\lambda\frac{1}{1-y}x}x^{m(T-t)-1}\re^{-x}\rd x \nonumber \\ &=&
\frac{1}{\Gamma[m(T-t)]}\int^{\infty}_0\
x^{mT-1}\re^{-\frac{x}{1-y}}\re^{{\rm i}\lambda\frac{x}{1-y}}\rd x
\nonumber \\ &=& \frac{(1-y)^{mT}}{\Gamma[m(T-t)]}
\int^{\infty}_0\,u^{mT-1} \re^{-u}\re^{{\rm i}\lambda u}\rd u
\nonumber \\ &=& \frac{(1-y)^{mT}}{\Gamma[m(T-t)]}\
\frac{\Gamma[mT]}{(1-\ri\lambda)^{mT}}, \label{eq:31}
\end{eqnarray}
where we have made use of (\ref{eq:1.4}) to deduce that the
characteristic function of $\gamma_T$ is
$1/(1-\ri\lambda)^{mT}$. Substituting (\ref{eq:31}) into
(\ref{proof2-eq1}) we obtain
\begin{eqnarray}
f(y,z)=\frac{\Gamma[mT]}{\Gamma[mt]\Gamma[m(T-t)]}\
y^{mt-1}(1-y)^{m(T-t)-1}\frac{1}{2\pi}
\int^{\infty}_{-\infty}\frac{1}{(1-\ri\lambda)^{mT}}\ \re^{-{\rm
i}\lambda z}\rd\lambda,
\end{eqnarray}
and hence
\begin{eqnarray}
f(y,z)=\frac{y^{mt-1}(1-y)^{m(T-t)-1}}{\textrm{B}[mt,m(T-t)]}\
\frac{z^{mT-1}\re^{-z}}{\Gamma[mT]}
\end{eqnarray}
for $0<y<1$ and $z>0$. Thus we see that the joint density for
$\gamma_t/\gamma_T$ and $\gamma_T$ factorises into the product of
a beta density for $\gamma_t/\gamma_T$ and a gamma density for
$\gamma_T$, as desired. \hfill$\Box$ \vspace{0.4cm}

The result of Proposition~\ref{prop:2} is a special case
of the following more general result:

\begin{prop} \label{prop:3}
Let $\{\gamma_t\}_{0\le t<\infty}$ be a standard gamma process.
Then for $T\ge u\ge t\ge 0$ the random variables
$(\gamma_u-\gamma_t)/(\gamma_T-\gamma_t)$ and $\gamma_T-\gamma_t$
are independent.
\end{prop}

Clearly, Proposition~\ref{prop:2} follows as a special case of
Proposition~\ref{prop:3}. The following lemma is a classical result
(Lukacs 1955, Yeo \& Milne 1991) which can be used as the basis of a
proof of Proposition~\ref{prop:3}.

\begin{lem} \label{lem:1}
Let $A$ and $B$ be independent gamma-distributed random variables
with parameters $p$ and $q$, respectively. Then $A/(A+B)$ and $A+B$
are independent, $A/(A+B)$ has a {\rm beta}$(p,q)$ distribution, and
$A+B$ has a {\rm gamma}$(p+q)$ distribution.
\end{lem}

\noindent{\bf Proof}. For independence it suffices to show
that the joint Laplace transform of $A/(A+B)$
and $A+B$ factorises. In particular, for positive $\alpha,\beta$
we have
\begin{eqnarray}
{\mathbb E}\left[ \re^{-\alpha A/(A+B)-\beta(A+B)}\right] =
\int_0^\infty \! \int_0^\infty \frac{a^{p-1}\re^{-a}}{\Gamma[p]}
\, \frac{b^{q-1}\re^{-b}}{\Gamma[q]}\, \re^{-\alpha a/(a+b)- \beta
(a+b)} \rd a \,\rd b.
\end{eqnarray}
Setting $x=a/(a+b)$ and $y=a+b$, we have $a=xy$ and $b=(1-x)y$,
and hence $\rd a\,\rd b=y\,\rd x\,\rd y$. We see that
\begin{eqnarray}
{\mathbb E}\left[ \re^{-\alpha A/(A+B)-\beta(A+B)}\right] &=&
\int_0^1 \frac{x^{p-1}(1-x)^{q-1}}{{\rm B}[p,q]}\, \re^{-\alpha x}
\rd x \int_0^\infty \frac{y^{p+q-1}\re^{-y}}{\Gamma[p+q]}\,
\re^{-\beta y} \rd y \nonumber \\ &=& {\mathbb E}\left[
\re^{-\alpha A/(A+B)}\right] {\mathbb E}\left[ \re^{-\beta(A+B)}
\right].
\end{eqnarray}
It follows that $A/(A+B)$ and $A+B$ are independent and have the
distributions stated.  \hfill$\Box$ \vspace{0.4cm}

The proof of Proposition~\ref{prop:3} follows if we set
$A=\gamma_u-\gamma_t$ and $B=\gamma_T-\gamma_u$. A proof of
Proposition~\ref{prop:2} is obtained if we set $A=\gamma_t$ and
$B=\gamma_T-\gamma_t$.

\section{Valuation of aggregate claims}
\label{sec:5}

Our objective is to calculate the value at $t$ of a contract that
pays $X_T$ at $T$. We assume that $X_T$ is strictly positive and
integrable. For simplicity of exposition, in this section we take
$X_T$ to be a continuous random variable; the adjustments required
for the more general situation are straightforward. We assume that
the default-free interest rate system is deterministic, that
${\mathbb Q}$ is the risk-neutral measure, and that the market
filtration is generated by an aggregate claims process
$\{\xi_t\}_{0\le t\le T}$ of the form $\xi_t=X_T\gamma_{tT}$, where
$\{\gamma_{tT}\}$ is a standard gamma bridge under ${\mathbb Q}$,
with parameter $m$, which we take to be independent of $X_T$. The
value $S_t$ of the contract at $t\leq T$ is given by
$S_t=P_{tT}{\mathbb E}\left[ X_T\, \vert \,{\mathcal F}_t\right]$,
where ${\mathcal F}_t=\sigma\left(\{\xi_s\}_{0\le s\le t}\right)$.

\begin{prop} \label{prop:4}
The aggregate claims process $\{\xi_t\}_{0\le t\le T}$ has the
Markov property.
\end{prop}

\noindent{\bf Proof}. For the Markov property we must verify that
\begin{eqnarray}
{\mathbb Q}\left[\xi_t<a\,\vert\,{\mathcal F}_s\right]={\mathbb
Q}\left[\xi_t<a\,\vert\,\xi_s\right]
\end{eqnarray}
for all $s,t$ such that $0\leq s\leq t\leq T$. It suffices to
establish that
\begin{eqnarray}
{\mathbb Q}\left[\xi_t<a\,\vert\,\xi_s,\xi_{s_1},\xi_{s_2},
\ldots, \xi_{s_n}\right]={\mathbb
Q}\left[\xi_t<a\,\vert\,\xi_s\right]
\end{eqnarray}
for all $t\ge s\ge s_1\ge s_2\ge\cdots\ge s_n$, and for all $n\ge
1$. We use the representation $\{\gamma_{tT}\}
=\{\gamma_t/\gamma_T\}$, where $\{\gamma_t\}$ is a standard gamma
process with rate $m$. Then we have
\begin{eqnarray}
{\mathbb
Q}\left[\xi_t<a\,\vert\,\xi_s,\xi_{s_1},\xi_{s_2},\ldots\right]
&=&{\mathbb Q}\left[\xi_t<a\,\bigg\vert\,X_T
\frac{\gamma_s}{\gamma_T},
X_T\frac{\gamma_{s_1}}{\gamma_T},X_T\frac{\gamma_{s_2}}{\gamma_T},
\ldots\right]\nonumber\\
&=&{\mathbb Q}\left[\xi_t<a\,\bigg\vert\,X_T
\frac{\gamma_s}{\gamma_T},\frac{\gamma_{s_1}}{\gamma_s},
\frac{\gamma_{s_2}}{\gamma_{s_1}},\ldots\right].
\end{eqnarray}
But $\gamma_{s_1}/\gamma_{s},\gamma_{s_2}/\gamma_{s_1},\ldots$ are
independent of $\xi_t$ and $\xi_s$, which gives us the desired
result. \hfill$\Box$ \\

By virtue of the fact that $\{\xi_t\}$ has the Markov property and
that $X_T$ is ${\mathcal F}_T$-measurable we are able to simplify
the expression for $S_t$ so that it takes the form
\begin{eqnarray}\label{valuation}
S_t=P_{tT}{\mathbb E}\left[X_T\,\vert\,\xi_t\right].
\end{eqnarray}
The conditional expectation appearing here can be carried out in
closed form, leading to the following pricing formula:

\begin{prop} \label{prop:5}
The value $S_t$ at time $t<T$ of the aggregate claim that pays the
continuous random variable $X_T>0$ at time $T$ is given by
\begin{eqnarray}
S_t=P_{tT}\frac{\int^{\infty}_{\xi_t}p(x)x^{2-mT}
(x-\xi_t)^{m(T-t)-1}\,\rd x}{\int^{\infty}_{\xi_t}p(x)x^{1-mT}
(x-\xi_t)^{m(T-t)-1}\,\rd x}, \label{gammaAsset}
\end{eqnarray}
where $\{p(x)\}_{0<x<\infty}$ is the probability density
of $X_T$.
\end{prop}

\noindent{\bf Proof}. The conditional expectation
(\ref{valuation}) can be written in the form
\begin{eqnarray}
{\mathbb E}\left[X_T\,\vert\, \xi_t\right]=\int^{\infty}_0
x\pi_t(x)\rd x,
\end{eqnarray}
where $\left\{\pi_t(x)\right\}$ is the conditional density process
for $X_T$, which by virtue of the Markov property of $\{\xi_t\}$ is
given by
\begin{eqnarray}
\pi_t(x)=\frac{\rd}{\rd x}\,{\mathbb Q}\left[X_T\le x\,\vert\,
\xi_t\right].
\end{eqnarray}
We can compute $\pi_t(x)$ by use of the following form of the
Bayes formula:
\begin{eqnarray}
\pi_t(x)=\frac{p(x)\rho\left(\xi_t\,\vert\,X_T=x\right)}
{\int^{\infty}_0 p(x)\rho(\xi_t\,\vert\,X_T=x)\rd x},
\end{eqnarray}
where $\rho\left(\xi_t\,\vert\,X_T=x\right)$ is the conditional
density for $\xi_t$, valued at $\xi_t$. Specifically, we have
\begin{eqnarray}
\rho\left(\xi\,\vert\,X_T=x\right)&=&\frac{\rd}{\rd \xi}\,
{\mathbb Q}\left[\xi_t\le \xi\,\vert\,X_T=x\right]\nonumber\\
&=&\frac{\rd}{\rd \xi}\,{\mathbb Q}\left[X_T\gamma_{tT}\le \xi\,
\vert\,X_T=x\right] \nonumber\\ &=&\frac{\rd}{\rd \xi}\,{\mathbb Q}
\left[\gamma_{tT}\leq \frac{\xi}{x}\right].
\end{eqnarray}
Therefore, writing $\{f(y)\}_{0<y<1}$ for the density function of
the random variable $\gamma_{tT}$ we find
\begin{eqnarray}
\rho\left(\xi\,\vert\,X_T=x\right) &=&\frac{\rd}{\rd \xi}
\int_0^{\xi/x} f(y)\rd y\nonumber\\ &=&\frac{1}{x}\
f\left(\frac{\xi}{x}\right).
\end{eqnarray}
Hence by Proposition~\ref{prop:1} we have
\begin{eqnarray}
\rho\left(\xi\,\vert\,X_T=x\right)&=&\frac{1}{x}\, {\mathds
1}_{\{x>\xi\}}\frac{(\xi/x)^{mt-1}(1-\xi/x)^{m(T-t)-1}}
{\textrm{B}\left[mt,m(T-t)\right]}\nonumber\\ &=&{\mathds
1}_{\{x>\xi\}} \xi^{mt-1}\frac{x^{1-mT}(x-\xi)^{m(T-t)-1}}
{\textrm{B}\left[mt,m(T-t)\right]}.
\end{eqnarray}
The conditional probability density function $\{\pi_t(x)\}_{0\leq
t<T,\, x>0}$ for $X_T$ is thus given by
\begin{eqnarray}
\pi_t(x)={\mathds 1}_{\{x>\xi_t\}} \frac{p(x)x^{1-mT}
(x-\xi_t)^{m(T-t)-1}} {\int^{\infty}_{\xi_t}p(x)x^{1-mT}
(x-\xi_t)^{m(T-t)-1} {\rm d} x}, \label{eq:5.55}
\end{eqnarray}
from which the desired result (\ref{gammaAsset}) follows at once.
\hfill$\Box$\\

With these results at hand we are also in a position to price a
simple stop-loss reinsurance policy. For such a policy the value
process is given by (2), and hence we have
\begin{eqnarray}
C_{tT} &=& P_{tT} \int^{\infty}_0 (x-K)^+\pi_t(x)\rd x \nonumber \\
&=& P_{tT}\frac{\int^{\infty}_{\xi_t}(x-K)^+p(x)x^{1-mT}
(x-\xi_t)^{m(T-t)-1}\,\rd x}{\int^{\infty}_{\xi_t}p(x)x^{1-mT}
(x-\xi_t)^{m(T-t)-1}\,\rd x} .
\end{eqnarray}
It should be evident that once a time $t$ has been reached such that
$\xi_t \ge K$, then $C_{uT}=P_{uT}(S_t-K)$ for all $u$ such that
$t\leq u\leq T$. In other words, once a sufficient number of claims
have accumulated the option is sure to expire in-the-money.

\section{Valuation of general reinsurance contracts}
\label{sec:6}

In the previous section we showed how one works out the reserve
process for an aggregate claim that pays $X_T$ at $T$, and we were
also able to determine the value process of a stop-loss contract
that pays $(X_T-K)^+$ at $T$. In this section we consider the more
general situation of a contract that at a fixed time $t<T$ allows
the policy holder the option of commuting the claim $X_T$ in
exchange for a pre-fixed settlement $K$. Let us write $C_{0t}$ for
the value at time $0$ of such an option; then clearly we have
\begin{eqnarray}
C_{0t}=P_{0t} {\mathbb E} \left[ (S_t-K)^+\right],
\end{eqnarray}
where $S_t$ is the value at $t$ of the claim that pays $X_T$ at $T$.
With reference to Proposition~\ref{prop:5}, it will be useful to
introduce a function $S(t,y)$ for $0\leq t<T$ and $y\geq 0$ by
setting
\begin{eqnarray}
S(t,y)=P_{t T} \frac{\int_y^\infty p(x) x^{2-mT} (x-y)^{m(T-t)-1}
\rd x} {\int_y^\infty p(x)x^{1-mT} (x-y)^{m(T-t)-1}\rd x}.
\label{eq:zz2}
\end{eqnarray}
Then the value of the claim is given by $S_t=S(t,\xi_t)$, and the
value of the option can be written in the form
\begin{eqnarray}
C_{0t}=P_{0t}{\mathbb E} \left[\left(S(t,\xi_t)-K\right)^+\right].
\label{eq:zz1}
\end{eqnarray}
Since the payout of the option is a function of $\xi_t$, one way of
working out the expectation in (\ref{eq:zz1}) is to obtain an
expression for the price $A_{0t}(y)$ of an Arrow-Debreu security
that pays $\delta(\xi_t-y)$ at $t$, where $y\ge0$ is a parameter.
Thus we have
\begin{eqnarray}
A_{0t}(y)=P_{0t}{\mathbb E}[\delta(\xi_t-y)],
\end{eqnarray}
and for the option we can write
\begin{eqnarray}
C_{0t} = \int_0^\infty A_{0t}(y) \left[ S(t,y)-K\right]^+ \rd y.
\label{eq:zz4}
\end{eqnarray}
We shall calculate
$A_{0t}(y)$ and use the result to determine the expectation
(\ref{eq:zz1}). We state the result of this calculation first,
the proof of which is given at the end of this section.

\begin{prop} \label{prop:6} The price $A_{0t}(y)$ at time $0$ of
an Arrow-Debrue security that pays $\delta(\xi_t-y)$ at $t$ is given
by
\begin{eqnarray}
A_{0t}(y)=P_{0t} \frac{y^{mt-1}}{\textrm{B} [mt,m(T-t)]}
\int_{y}^{\infty}p(x)\, x^{1-mT} (x-y)^{m(T-t)-1}\rd x.
\label{eq:54}
\end{eqnarray}
\end{prop}

By comparing (\ref{eq:zz2}) and (\ref{eq:54}) we observe that the
integral term in (\ref{eq:54}) cancels with the denominator in the
expression for $S(t,y)$. After some rearrangement we thus obtain
\begin{eqnarray}
C_{0t}&=&\int_0^\infty \frac{P_{0t}\,y^{mt-1}}{\textrm{B}
[mt,m(T-t)]} \left[\int_y^\infty \!p(x) \left( xP_{tT}-K\right)
x^{1-mT} (x-y)^{m(T-t)-1} \rd x\right]^+\rd y. \label{eq:57}
\end{eqnarray}

We are now left with the task of finding the critical values at
which the argument of the max-function in the integrand of
(\ref{eq:57}) vanishes. Suppose that $S(t,y)$ is monotonic in $y$;
then there is at most a single critical value $y^*$, obtained by
solving the following equation:
\begin{eqnarray}
\int^{\infty}_{y^*}p(x)\left( xP_{tT}-K\right)
\,x^{1-mT}(x-y^*)^{m(T-t)-1}\rd x = 0. \label{eq:60}
\end{eqnarray}
The lower limit of the outer integration in the expression for
$C_{0t}$ above can then be changed, and we have
\begin{eqnarray}
C_{0t}&=&\int_{y^*}^\infty \frac{P_{0t}\,y^{mt-1}}{\textrm{B}
[mt,m(T-t)]} \left[\int_{y}^\infty \!p(x) \left( xP_{tT}-K\right)
x^{1-mT} (x-y)^{m(T-t)-1} \rd x\right]\rd y. \label{eq:57a}
\end{eqnarray}
This expression simplifies further if we swap the order of
integration as follows:
\begin{eqnarray}
C_{0t}&=&\frac{P_{0t}}{\textrm{B}[mt,m(T-t)]} \int_{y^*}^\infty
\int_{y^*}^x \!p(x) \left( xP_{tT}-K\right) y^{mt-1}x^{1-mT}
(x-y)^{m(T-t)-1} \rd y\, \rd x. \label{eq:57b}
\end{eqnarray}
Making the substitution $y=xz$ we then obtain
\begin{eqnarray}
C_{0t}=\frac{P_{0t}}{\textrm{B}[mt,m(T-t)]} \int^{\infty}_{y^*} p(x)
\left( xP_{tT}-K\right) \int_{y^*/x}^1 z^{mt-1} (1-z)^{m(T-t)-1} \rd
z\,\rd x. \label{eq:76}
\end{eqnarray}
Let us now introduce the complementary beta distribution function
${\mathcal B}(u)$ with parameters $mt$ and $m(T-t)$ by the
expression:
\begin{eqnarray}
{\mathcal B}(u) = \frac{\int_{u}^{1} z^{mt-1} (1-z)^{m(T-t)-1} \rd
z}{\int_{0}^{1} z^{mt-1} (1-z)^{m(T-t)-1} \rd z} . \label{eq:93}
\end{eqnarray}
We call this the ``complementary'' distribution because the
integration ranges from $u$ to $1$.
Clearly, the denominator in (\ref{eq:93}) is
$B[mt,m(T-t)]$. We thus find that the integration over the variable
$z$ in (\ref{eq:76}) combines with the factor $B[mt,m(T-t)]$
appearing of that expression to give a cumulative
beta distribution function, and for the option price we have
\begin{eqnarray}
C_{0t}=P_{0t}\int^{\infty}_{y^*} p(x) \left(x\,P_{tT}-K\right)
{\mathcal B}(y^*/x)\, \rd x. \label{eq:zz3}
\end{eqnarray}

We remark, incidentally, that a sufficient condition for $S(t,y)$ to
be monotonic in $y$ for fixed $t$ is $m(T-t)>1$. To see this, we
differentiate $S(t,y)$ with respect to $y$, assuming the stated
condition, and after some rearrangement we obtain
\begin{eqnarray}
\frac{\partial S(t,y)}{\partial y}=P_{tT}\left[ m(T-t)-1\right]
\left( \frac{\int_y^\infty p(x) \alpha^2(x)\rd x \int_y^\infty p(x)
\beta^2(x)\rd x}{\left(\int_y^\infty p(x) \alpha(x)\beta(x)\rd x
\right)^2}-1 \right), \label{eq:59}
\end{eqnarray}
where $\alpha^2(x)=x^{1-mT}(x-y)^{m(T-t)}$ and $\beta^2(x)
=x^{1-mT}(x-y)^{m(T-t)-2}$. If $m(T-t)>1$, then the integrals exist,
and it follows on account of the Schwartz inequality that $\partial
S(t,y)/\partial y>0$. \vspace{0.2cm}

\noindent{\bf Proof of Proposition~\ref{prop:6}}. It suffices to
determine the expectation ${\mathbb E}[\delta(\xi_t-y)]$. By use of
the Fourier representation (\ref{deltarep})
we can write
\begin{eqnarray}
{\mathbb E}[\delta(\xi_t-y)] = \frac{1}{2\pi}
\int_{-\infty}^{\infty} \re^{-{\rm i}\lambda y} \, {\mathbb E}
\left[ \re^{{\rm i} \lambda \xi_t }\right] \rd\lambda .
\end{eqnarray}
Since $X_T$ and $\gamma_{t T}$ are independent, it follows from the
tower property that
\begin{eqnarray}
{\mathbb E} \left[ \re^{{\rm i}\lambda \xi_t} \right] &=& {\mathbb
E} \Big[{\mathbb E} \left[\re^{{\rm i} \lambda X_T \gamma_{t T}}
\vert X_T\right]\Big] \nonumber\\ &=& \int_0^\infty p(x)\, {\mathbb
E} \left[\re^{{\rm i}\lambda x \gamma_{tT}} \right] \rd x \nonumber
\\ &=& \int_0^\infty p(x)\,\phi(\lambda x) \,\rd x,
\end{eqnarray}
where $\phi(\nu)={\mathbb E}\left[\re^{{\rm i} \nu \gamma_{tT}}
\right]$ is the characteristic function of $\gamma_{tT}$. We deduce
that
\begin{eqnarray}
{\mathbb E}[\delta(\xi_t-y)]=\frac{1}{2\pi}\int_{-\infty}^{\infty}
\re^{-{\rm i}\lambda y}\int_{x=0}^\infty p(x)\,\phi(\lambda x)\, \rd
x\,\rd\lambda.
\end{eqnarray}
Thus, by interchanging the order of integration and using the fact
that the inverse Fourier transform of the characteristic function is
the density function we have
\begin{eqnarray}
{\mathbb E} [\delta(\xi_t-y)]&=&\int^{\infty}_{x=0}p(x)
\left[\frac{1}{2\pi}\int^{\infty}_{-\infty}\re^{-{\rm i}
\lambda y} \phi(\lambda x)\,\rd \lambda\right]\rd x\nonumber\\
&=&\int^{\infty}_{x=0}p(x)\frac{1}{x}\left[\frac{1}{2\pi}
\int^{\infty}_{-\infty}\re^{-{\rm i}\, \nu\, y/x} \phi(\nu)
\,\rd\nu\right]\rd x\nonumber\\ &=& \int_0^{\infty} p(x)
\frac{1}{x}f\left(\frac{y}{x}\right) \rd x, \label{eq:51}
\end{eqnarray}
where $f$ is the density function of $\gamma_{t T}$. Substituting
the expression (\ref{prop1}) for $f$ into (\ref{eq:51}) we find that
\begin{eqnarray}
{\mathbb E}\left[\delta(\xi_t-y)\right]&=&\int_0^{\infty} p(x) \,
\frac{1}{x}\,{\mathds 1}_{\{x>y\}} \frac{(y/x)^{mt-1}
(1-y/x)^{m(T-t)-1}}{\textrm{B} [mt,m(T-t)]}\,\rd x \nonumber \\ &=&
\frac{y^{mt-1}}{\textrm{B}[mt,m(T-t)]}\int_y^{\infty}
p(x)\,x^{1-mT}(x-y)^{m(T-t)-1}\rd x,
\end{eqnarray}
which verifies the claim. \hfill$\Box$ \vspace{0.4cm}

We remark that the price of the Arrow-Debrue security can be put in
the form
\begin{eqnarray}
A_{0t}(y) = P_{0t} \frac{\int_0^1 p(y/u)\, u^{mt-2}(1-u)^{m(T-t) -
1}\rd u}{\int_0^1 u^{mt-1}(1-u)^{m(T-t)-1}\rd u}, \label{eq:54-5}
\end{eqnarray}
by use of which the normalisation $\int_0^\infty A_{0t}(y)\rd y =
P_{0t}$ can be checked. It follows also from (\ref{eq:54-5}) that
the characteristic function $\Phi_\xi(\lambda)$ of $\xi_t$ is given
by the beta average of the characteristic function $\phi_X$ of
$X_T$:
\begin{eqnarray}
\Phi_\xi(\lambda) = \frac{\int_0^1 \phi_X(\lambda u)
u^{mt-1}(1-u)^{m(T-t) - 1}\rd u}{\int_0^1 u^{mt-1}
(1-u)^{m(T-t)-1}\rd u}. \label{eq:54-55}
\end{eqnarray}

\section{Discrete cash flows}
\label{sec:7}

Thus far we have considered the case for which the terminal cash
flow is a continuous random variable. In this section we consider
the example for which $X_T$ takes values in a discrete set
$\{x_i\}_{i=1,...,n}$. The corresponding \textit{a priori}
probabilities will be denoted $\{p_i\}$. The calculation presented
in Section~\ref{sec:5} holds and we obtain, instead of
(\ref{gammaAsset}), the following expression for the value process:
\begin{eqnarray}
S_t = P_{tT} \frac{\sum_i p_i x_i^{2-mT}(x_i-\xi_t)^{m(T-t)-1}
{\mathds 1}_{\{\xi_t< x_i\}}}{\sum_i p_i x_i^{1-mT} (x_i-
\xi_t)^{m(T-t)-1} {\mathds 1}_{\{\xi_t< x_i\}}}. \label{eq:9.1}
\end{eqnarray}
It is straightforward to verify that expression (\ref{eq:9.1})
converges to the correct terminal value as $t$ approaches $T$. To
see this, suppose that for some $\omega\in\Omega$ the value of $X_T$
is $x_k$. Then for that choice of $\omega$ we have
\begin{eqnarray}
S_t = P_{tT} \frac{\sum_{i} p_i x_i^{2-mT}(x_i-x_k\gamma_{tT}
)^{m(T-t)-1} {\mathds 1}_{\{x_i>x_k\gamma_{tT}\}}}{\sum_{i} p_i
x_i^{1-mT} (x_i-x_k\gamma_{tT} )^{m(T-t)-1} {\mathds
1}_{\{x_i>x_k\gamma_{tT}\}}}, \label{eq:9.3.0}
\end{eqnarray}
and hence, after some rearrangement,
\begin{eqnarray}
S_t &=& P_{tT} \frac{p_k x_k^{1-mt} + \sum_{i\neq k} p_i x_i^{2-mT}
\left(\frac{1-\gamma_{tT}}{x_i-x_k\gamma_{tT}}\right)^{1-m(T-t)}
{\mathds 1}_{\{x_i>x_k\gamma_{tT}\}}}{p_k x_k^{-mt}+\sum_{i\neq k}
p_i x_i^{1-mT} \left(\frac{1-\gamma_{tT}}{x_i-x_k\gamma_{tT}}
\right)^{1-m(T-t)} {\mathds 1}_{\{x_i>x_k\gamma_{tT}\}}} .
\label{eq:9.3}
\end{eqnarray}
It follows at once that $S_T=x_k$.

\begin{figure}
\begin{center}
  \includegraphics[scale=0.6]{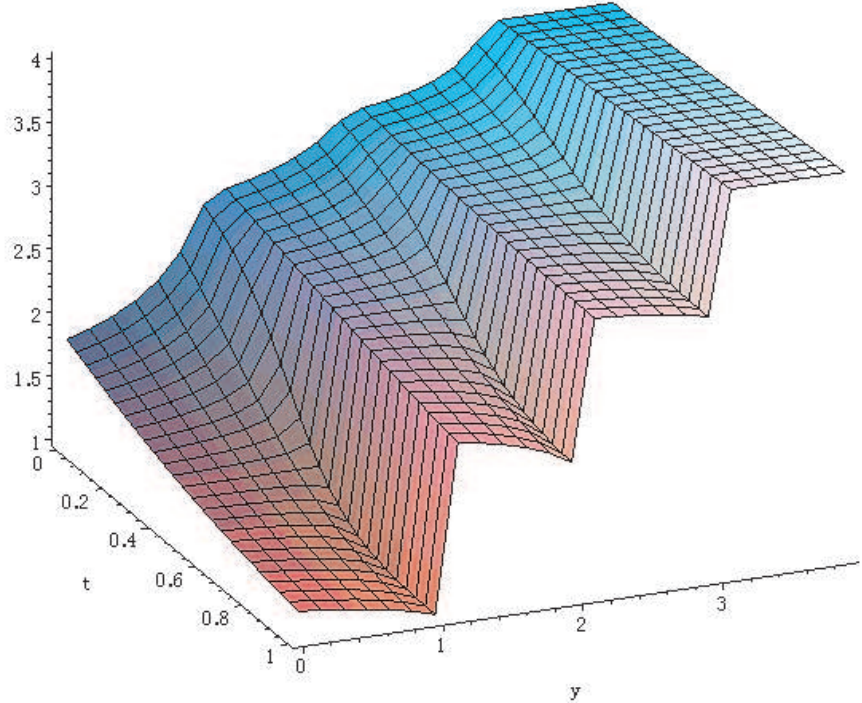}
  \caption{The value function $S(t,y)$ associated with the reserve
  price in the case of a
  discrete cash flow at time $T$
  taking four possible values. The parameters are
  chosen such that $\{x_1,x_2,x_3,x_4\}=\{1,2,3,4\}$,
  $\{p_1,p_2,p_3,p_4\}=\{0.5,0.2,0.2,0.1\}$, $m=2.0$, $r=5\%$, and
  $T=1$. For a given time $t$ the value function represents the
  reserve required if the aggregate claims amount to $y$.
  \label{fig:1}
  }
\end{center}
\end{figure}

We proceed now to value a reinsurance contract that pays $(S_t-K)^+$
at time $t$. For this purpose we need the price of an Arrow-Debreu
security with payoff $\delta(\xi_t-y)$ at $t$. In the discrete case
the Arrow-Debreu price is given by
\begin{eqnarray}
A_{0t}(y)=P_{0t} \frac{y^{mt-1}}{\textrm{B} [mt,m(T-t)]}
\sum_{i=0}^n p_i x_i^{1-mT} (x_i-y)^{m(T-t)-1} {\mathds
1}_{\{x_i>y\}}. \label{eq:87}
\end{eqnarray}
Substituting (\ref{eq:87}) and the function
\begin{eqnarray}
S(t,y) = P_{t T} \frac{\sum_i p_ix_i^{2-mT}(x_i-y)^{m(T-t)
-1}{\mathds 1}_{\{x_i>y\}}}{\sum_i p_i x_i^{1-mT} (x_i-y)^{m(T-t)-1}
{\mathds 1}_{\{x_i>y\}}} \label{eq:9.1a}
\end{eqnarray}
into (\ref{eq:zz4}) we obtain, after some rearrangement,
\begin{eqnarray}
C_{0t} = \frac{P_{0t}}{B[mt,m(T-t)]} \int\limits_0^\infty \!
y^{mt-1} \! \left[ \sum_{i=1}^n p_i x_i^{1-mT}(x_i-y)^{m(T-t)
-1}{\mathds 1}_{\{x_i>y\}}(P_{t T}x_i-K) \right]^+ \! \rd y.
\label{eq:90}
\end{eqnarray}
A discrete version of formula (\ref{eq:59}) shows that $S(t,y)$ is
increasing in $y$ if $m(T-t)>1$, and decreasing in $y$ for
$y\in(x_k, x_{k+1})$ for each $k=1,\ldots,n-1$ if $m(T-t)<1$. See
Figure~\ref{fig:1} for the typical behaviour of $S(t,y)$ when $X_T$
takes four possible values. For fixed $t$ there is at most a single
critical value $y=y^*$ for which $S(t,y)=K$, when $y\neq x_k$ for
all $k$. We thus have three scenarios to consider, namely: (I)
$S(t,y)$ is increasing in $y$ at $y=y^*$; (II) the critical value
$y^*$ is at $y=x_k$ for some $k$; and (III) $S(t,y)$ is decreasing
in $y$ at $y=y^*$.

We therefore analyse the price of the reinsurance contract in
these different scenarios. In case (I) the integrand in
(\ref{eq:90}) is nonzero when $y\in(y^*,\infty)$, and we have
\begin{eqnarray}
C_{0t} = \frac{P_{0t}}{B[mt,m(T-t)]} \sum_{i=1}^n p_i x_i^{1-mT}
(P_{t T}x_i-K) \int\limits_{y^*}^\infty \! y^{mt-1} (x_i-y)^{m(T-t)
-1}{\mathds 1}_{\{x_i>y\}} \rd y. \label{eq:91}
\end{eqnarray}
The $y$ integration in (\ref{eq:91}) can be carried out by observing
that for $x_i>y^*$ we have
\begin{eqnarray}
\int_{y^*}^\infty y^{mt-1}(x_i-y)^{m(T-t)-1}{\mathds
1}_{\{x_i>y\}}\rd y &=& x_i^{mT-2} \int_{y^*}^{x_i} \left(
\frac{y}{x_i}\right)^{mt-1} \left(1-\frac{y}{x_i} \right)^{m(T-t)-1}
\rd y \nonumber \\ &=& x_i^{mT-1} \int_{y^*/x_i}^{1} z^{mt-1}
(1-z)^{m(T-t)-1} \rd z, \label{eq:92}
\end{eqnarray}
where we have made the substitution $y=x_i z$. Therefore, the price
of the reinsurance contract can be expressed in terms of the
complementary beta distribution function with parameters $mt$ and
$m(T-t)$:
\begin{eqnarray}
C_{0t} = P_{0t} \sum_{i=1}^n {\mathds 1}_{\{x_i>y^*\}} p_i(P_{t
T}x_i-K) {\mathcal B} (y^*/x_i) . \label{eq:94}
\end{eqnarray}
If there is no critical value in the range $(x_k,x_{k+1})$, then
$y^*=x_k$ for some $k$. Hence the pricing formula in case (II) is
identical to the result obtained in (\ref{eq:94}), with $y^*=x_k$.
In case (III) there are two distinct regions for which the integrand
in (\ref{eq:90}) is nonzero. These are given by $y\in[x_k,y^*)$ and
$y \geq x_{k+1}$ for some $k$. Hence the pricing formula is similar
to that obtained in (\ref{eq:94}), except there are additional terms
arising from the integration over the range $[x_k,y^*)$.

As an example of a discrete cash flow we consider the
binary case where $X_T$ can take the values $x_0,x_1$. In this
situation the critical value $y^*<x_0$ can be worked out by
solving
\begin{eqnarray}
p_0(P_{t T}x_0-K)x_0^{1-mT}(x_0-y^*)^{m(T-t)-1} = p_1(K- P_{t T}x_1)
x_1^{1-mT}(x_1-y^*)^{m(T-t)-1}
\end{eqnarray}
for $y^*$. A short calculation shows that
\begin{eqnarray}
y^* = \frac{\theta x_1-x_0}{\theta-1}, \quad {\rm where} \quad
\theta = \left[ \frac{p_1(K-P_{t T}x_1)}{p_0(P_{t T}x_0-K)}
\left(\frac{x_1}{x_0}\right)^{1-mT}\right]^{\frac{1}{m(T-t)-1}} .
\end{eqnarray}
It follows that the price of a reinsurance contract in the case of a
binary payoff is given by
\begin{eqnarray}
C_{0t} = p_0(P_{0T}x_0-P_{0t}K) {\mathcal B}(y^*/x_0) + p_1(P_{0T}
x_1 - P_{0t}K){\mathcal B}(y^*/x_1) .
\end{eqnarray}

\section{Option price process}
\label{sec:8}

We generalise now the analysis of Section~\ref{sec:6} to derive an
expression for the price process of a call option on the value of
the reserve $S_t$ at time $t$ associated with the claim $X_T$. As
before, we let $K$ be the strike. Then the value of the option at
time $s\leq t$ is given by
\begin{eqnarray}
C_{st}=P_{st}{\mathbb E} \left[\left(S(t,\xi_t)-K\right)^+ | \xi_s
\right].
\end{eqnarray}
Once again we find it convenient to obtain first the price process
for the Arrow-Debreu security. This is on account of the relation
\begin{eqnarray}
C_{st} &=& P_{st}{\mathbb E}\left[\left. \int_{0}^{\infty}
\delta(\xi_t-y) \left(S(t,y)-K\right)^+ \rd y\right| \xi_s\right]
\nonumber \\ &=& P_{st}\int_{0}^{\infty} {\mathbb E}\left[
\delta(\xi_t-y)|\xi_s\right] \left(S(t,y)-K\right)^+ \rd y
\nonumber\\ &=& \int_{0}^{\infty} A_{st}(y) \left(S(t,y)-K\right)^+
\rd y, \label{COPPAD}
\end{eqnarray}
where $S(t,y)$ is defined as in (\ref{eq:zz2}), and
$\{A_{st}\}_{0\le s\le t\le T}$ is given by
\begin{eqnarray}
A_{st}(y)=P_{st}{\mathbb E} \left[ \delta(\xi_t-y)|\xi_s\right].
\label{ADPP}
\end{eqnarray}
By taking the conditional expectation we obtain the following
result:

\begin{prop} \label{prop:7}
The price process $\{A_{st}(y)\}_{0\le s\le t\le T}$ of the
Arrow-Debreu security that pays out $\delta(\xi_t-y)$ at $t$ is
given by
\begin{eqnarray}
A_{st}(y)=P_{st}\frac{{\mathds 1}_{\{y>\xi_s\}}
(y-\xi_s)^{m(t-s)-1}} {B[m(t-s),m(T-t)]}
\frac{\int^{\infty}_{y}p(x)\, x^{1-mT}(x-y)^{m(T-t)-1}\rd
x}{\int^{\infty}_{\xi_s} p(x)\,x^{1-mT}(x-\xi_s)^{m(T-s)-1} \rd x},
\label{eq:zz5}
\end{eqnarray}
where $y\ge \xi_s$ and $\{p(x)\}$ is the probability density of
$X_T$.
\end{prop}

This result is established later in this section. By substitution of
(\ref{eq:zz5}) in (\ref{COPPAD}) we see that the price process of
the option is given by
\begin{eqnarray}
C_{st} &=& \frac{P_{st}}{B[m(t-s),m(T-t)]\int^{\infty}_{\xi_s}
p(x)\,x^{1-mT}(x-\xi_s)^{m(T-s)-1}\rd x} \nonumber\\
&& \times \int^{\infty}_{y=\xi_s} (y-\xi_s)^{m(t-s)-1}
\left[\int^{\infty}_y p(x)x^{1-mT} (x-y)^{m(T-t)-1} (P_{tT}x-K) \rd
x\right]^+\rd y. \label{eq:zz6}
\end{eqnarray}
Assuming that there is only one critical value $y^*$ that solves
(\ref{eq:60}), we find that the integration over $y$ in
(\ref{eq:zz6}) vanishes for $y$ smaller than $y^*$. In this case, we
can lift the max-function in the integrand, and by interchanging the
order of integration we obtain
\begin{eqnarray}
C_{st} &=& \frac{P_{st}}{B[m(t-s),m(T-t)]
\int^{\infty}_{\xi_s}p(x)\,x^{1-mT}(x-\xi_s)^{m(T-s)-1}\rd x}
\nonumber \\ && \times P_{tT}\int^{\infty}_{x=y^*} p(x) x^{1-mT}
(P_{tT}x-K) \int^{x}_{y=y^*} (y-\xi_s)^{m(t-s)-1}(x-y)^{m(T-t)-1}
\,\rd y\,\rd x. \label{eq:zz8}
\end{eqnarray}
Let us analyse the $y$ integration. Making the substitution
$y=z+\xi_s$ we find that
\begin{eqnarray}
\int^{x}_{y=y^*}(y-\xi_s)^{m(t-s)-1}(x-y)^{m(T-t)-1}\,\rd
y=\int^{x-\xi_s}_{z=y-\xi_s}z^{m(t-s)-1}(x-\xi_s-z)^{m(T-t)-1}\rd z.
\end{eqnarray}
A further change of variable obtained by setting $w=z/(x-\xi_s)$
gives
\begin{eqnarray}
&& \int^{x-\xi_s}_{z=y-\xi_s}z^{m(t-s)-1}(x-\xi_s-z)^{m(T-t)-1}\rd z
\nonumber \\ && \qquad \qquad = (x-\xi_s)^{m(T-s)-1}
\int_{w=\frac{y^*-\xi_s}{x-\xi_s}}^1 w^{m(t-s)-1}(1-w)^{m(T-t)-1}
\rd w. \label{eq:zz9}
\end{eqnarray}
We see that together with the beta function in the denominator of
(\ref{eq:zz8}) the integral term in the right side of (\ref{eq:zz9})
gives rise to a complementary beta distribution function. Therefore,
the call price can be written in the form
\begin{eqnarray}
C_{st} = P_{st} \int^{\infty}_{x=y^*} \frac{p(x)x^{1-mT}
(x-\xi_s)^{m(T-s)-1}}{\int^{\infty}_{\xi_s}p(x)\,x^{1-mT}
(x-\xi_s)^{m(T-s)-1}\rd x} \,(P_{tT}x-K)\,
\mathcal{B}\left(\frac{y^*-\xi_s} {x-\xi_s}\right) \rd x .
\end{eqnarray}
Finally, we observe that the quotient in the integrand is the
conditional density $\pi_s(x)$. The call price at time
$s\leq t$ thus reduces to the following expression:
\begin{eqnarray}
C_{st}=P_{st}\int^{\infty}_{x=y*}\pi_s(x)(xP_{tT}-K)\,\mathcal{B}
\left(\frac{y^*-\xi_s}{x-\xi_s}\right)\,\rd x. \label{eq:zz10}
\end{eqnarray}
As in the case of the initial price of the option, the range of
integration in (\ref{eq:zz10}) must be modified appropriately if
there is more than one critical value for which (\ref{eq:60}) is
satisfied. We now proceed to derive the expression for the
Arrow-Debreu price process. \vspace{0.4cm}

\noindent{\bf Proof of Proposition~\ref{prop:7}}. By use of the
Fourier representation (\ref{deltarep}) we have
\begin{eqnarray}
{\mathbb E}\left[ \delta(\xi_t-y)|\xi_s\right] =\frac{1}{2\pi}
\int^{\infty}_{-\infty}\re^{-{\rm i}\lambda y}\,{\mathbb E}
\left[\left.\re^{{\rm i}\lambda\xi_t}\right| \xi_s\right]\rd\lambda.
\label{FourierADPP}
\end{eqnarray}
To determine the conditional expectation ${\mathbb E} \re^{{\rm
i}\lambda\xi_t}| \xi_s]$ the following result is handy:

\begin{lem} \label{lem:2}
Let $\{\xi_t\}_{0\le t\le T}$ be given by $\xi_t=X_T \gamma_{tT}$,
where $\{\gamma_{tT}\}$ is a gamma bridge and $X_T$ is an
independent positive random variable. Then for fixed $s$ such that
$0\le s\le t\le T$ we have
\begin{eqnarray}
\xi_t=\xi_s+Z_T \delta_{tT}, \label{eq:8.1}
\end{eqnarray}
where $Z_T=(1-\gamma_{sT})X_T$, and where the process
$\{\delta_{tT}\}_{s\le t\le T}$, defined by
\begin{eqnarray}
\delta_{tT}=\frac{\gamma_{tT}-\gamma_{sT}}{1-\gamma_{sT}},
\end{eqnarray}
is a gamma bridge over the interval $t\in[s,T]$ and is independent
of $\xi_s$ and $Z_T$.
\end{lem}

\noindent By use of (\ref{eq:8.1}) and the tower property we find
that
\begin{eqnarray}
{\mathbb E}\left[\left. \re^{{\rm i}\lambda\xi_t}\right| \xi_s
\right] &=& {\mathbb E}\left[\left.\re^{{\rm i}\lambda(\xi_s+Z_T
\delta_{tT})}\right| \xi_s\right] \nonumber\\ &=& \re^{{\rm i}
\lambda\xi_s} {\mathbb E} \left[\left. \re^{{\rm i} \lambda Z_T
\delta_{tT}}\right| \xi_s\right] \nonumber \\ &=& \re^{{\rm i}
\lambda\xi_s}{\mathbb E}\left[\left.{\mathbb E} \left[ \re^{{\rm i}
\lambda Z_T \delta_{tT}}\big| \xi_s, \delta_{tT} \right] \right|
\xi_s\right].
\end{eqnarray}
Since $Z_T=X_T-\xi_s$, and since $\{\delta_{tT}\}$ is independent of
$\xi_s$ and $X_T$, the inner expectation can be carried out
explicitly by use of the conditional density for $X_T$, and we
obtain
\begin{eqnarray}
{\mathbb E}\left[\left.\re^{{\rm i}\lambda\xi_t}\right|\xi_s\right]
=\re^{{\rm i}\lambda\xi_s}{\mathbb E}\left[\left.
\int^{\infty}_{x=\xi_s} \re^{{\rm i}\lambda(x-\xi_s) \delta_{tT}}
\pi_s(x)\rd x \right| \xi_s\right]. \label{eq:zz12}
\end{eqnarray}
By substituting (\ref{eq:zz12}) in (\ref{FourierADPP}) we deduce
that
\begin{eqnarray}
{\mathbb E}\left[\delta(\xi_t-y)|\xi_s\right] &=&
\frac{1}{2\pi}\int^{\infty}_{-\infty}\re^{-{\rm i}\lambda
(y-\xi_s)}\int^{\infty}_{x=\xi_s}\Phi_{\delta}[\lambda(x-
\xi_s)]\,\pi_s(x)\,\rd x\,\rd \lambda \nonumber \\ &=&
\int^{\infty}_{x=\xi_s}\left(\frac{1}{2\pi}\int^{\infty}_{-\infty}
\re^{-{\rm i}\lambda (y-\xi_s)} \,\Phi_{\delta}[\lambda(x-\xi_s)]
\,\rd\lambda\right)\pi_s(x)\,\rd x,
\end{eqnarray}
where $\Phi_{\delta}$ is the characteristic function for
$\delta_{tT}$. By use of the
substitution $z=\lambda(x-\xi_s)$ we then find that
\begin{eqnarray}
{\mathbb E}\left[\delta(\xi_t-y)|\xi_s\right] &=&
\int^{\infty}_{x=\xi_s} \left( \frac{1}{2\pi}
\int^{\infty}_{-\infty} \re^{-{\rm i} \frac{y-\xi_s}{x-\xi_s}
\,z}\,\Phi_{\delta}(z)\rd z\right)\frac{1}{x-\xi_s}\,\pi_s(x)\,\rd x
\nonumber \\ &=& \int^{\infty}_{x=\xi_s} \frac{\pi_s(x)}{x-\xi_s}
\,f_{\delta} \left(\frac{y-\xi_s}{x-\xi_s}\right)\rd x,
\end{eqnarray}
where $f_{\delta}$ is the probability density of $\delta_{tT}$.
Since $\delta_{tT}$ is beta distributed with parameters $m(t-s)$ and
$m(T-t)$, we deduce, after some rearrangement, the expression
obtained in (\ref{eq:zz5}) for the Arrow-Debreu price. \hfill$\Box$
\vspace{0.4cm}

\noindent {\bf Proof of Lemma~\ref{lem:2}}. The decomposition
(\ref{eq:8.1}) can be verified by direct calculation if one sets
$\{\gamma_{tT}\}= \{\gamma_t/\gamma_T\}$, where $\{\gamma_t\}$ is a
standard gamma process. To see that $\{\delta_{tT}\}_{s\le t\le T}$
is, for fixed $s$, a gamma bridge over $[s,T]$ it suffices to note
that $\delta_{tT}=(\gamma_t -\gamma_s)/(\gamma_T-\gamma_s)$ and that
$\{\gamma_t- \gamma_s \}_{s\leq t<\infty}$ is a gamma process. In
particular, we observe that the independent increments property
holds, and that $\gamma_t-\gamma_s$ is gamma distributed with mean
$m(t-s)$. Finally, to see that $\{\delta_{tT}\}$ is independent of
$\xi_s$ and $Z_T$ it suffices to show that $\delta_{tT}$, $\gamma_s$
and $\gamma_T-\gamma_s$ are independent. We have:
\begin{eqnarray}
\mathbb{Q}\big(\{\delta_{tT}<a\}\cap\{\gamma_T-\gamma_s<b\}\cap
\{\gamma_s<c\}\big)&=& {\mathbb E}\!\left[{\mathds 1}_{\{
\delta_{tT}<a\}} {\mathds 1}_{\{ \gamma_T-\gamma_s<b\}} {\mathds
1}_{ \{\gamma_s<c\}}\right]\nonumber\\ &=& {\mathbb E} \! \left[
{\mathbb E}\! \left[{\mathds 1}_{\{\delta_{tT}<a\}}{\mathds 1}_{
\{\gamma_T-\gamma_s<b\}} {\mathds 1}_{ \{ \gamma_s<c\}}
\,\vert\,\gamma_s\right]\right]\nonumber\\ &=& {\mathbb E}\! \left[
{\mathbb E}\!\left[{\mathds 1}_{\{\delta_{tT}<a\}}{\mathds 1}_{\{
\gamma_T-\gamma_s<b\}} \,\vert\,\gamma_s\right] {\mathds 1}_{\{
\gamma_s<c\}}\right]\nonumber\\ &=&{\mathbb E}\!\left[{\mathbb E}\!
\left[{\mathds 1}_{\{\delta_{tT}<a\}}{\mathds 1}_{\{
\gamma_T-\gamma_s<b\}}\right]{\mathds 1}_{\{\gamma_s<c\}} \right]
\nonumber\\ &=&{\mathbb E}\!\left[{\mathds 1}_{\{\delta_{tT}<a\}}
{\mathds 1}_{\{\gamma_T-\gamma_s<b\}} \right] {\mathbb E}\! \left[
{\mathds 1}_{\{\gamma_s<c\}}\right]\nonumber\\ &=&{\mathbb E}\!
\left[ {\mathds 1}_{\{\delta_{tT}<a\}}\right]{\mathbb E}\!
\left[{\mathds 1}_{ \{\gamma_T-\gamma_s<b\}}\right]{\mathbb E}\!
\left[{\mathds 1}_{\{ \gamma_s<c\}}\right].
\end{eqnarray}
In going from the fourth to the fifth line we have used the fact
that $\gamma_s$ is independent of $\delta_{tT}$ and
$\gamma_T-\gamma_s$, which can be checked by use of the independent
increments property of $\{\gamma_t\}$. In going from the sixth to
seventh line we have used Lemma~\ref{lem:1} together with the fact
that we can write $\delta_{tT}=B/(A+B)$ and $\gamma_T-\gamma_s=A+B$,
with $A=\gamma_T-\gamma_t$ and $B=\gamma_t-\gamma_s$, from which it
follows that $\delta_{tT}$ and $\gamma_T-\gamma_s$ are independent.
\hfill$\Box$ \vspace{0.4cm}

The result of Lemma~\ref{lem:2} leads to the following observation
concerning the model calibration. Suppose that the aggregate claims
process is given, and that we reinitialise the model at some
specified intermediate time. We would like the dynamics of the model
moving forward from that intermediate time to be consistently
represented by an aggregate claims process of the same type. Indeed,
it follows from Lemma~\ref{lem:2} that the process $\{\eta_t\}_{s\le
t\le T}$ defined by
\begin{eqnarray}
\eta_t=Z_T \delta_{tT}
\end{eqnarray}
is an aggregate claims process spanning the time interval $[s,T]$.
The random variable $Z_T$ can be thought of as representing the
information about $X_T$ that is ``not yet revealed'' at time $s$.
The idea is that at time $s$ the value of $\xi_s$ is known, and the
``new'' gains process $\{\eta_t\}_{s\le t\le T}$ begins to reveal
the value of $Z_T$ in such a way that $\eta_s=0$ and $\eta_T=Z_T$.

Alternatively, at time $s$ we can use the knowledge of $\xi_s$ to
compute the ``new'' \textit{a priori} density for $X_T$. Thus, at
time $s$ the \textit{a priori} density $p(x)$ for $X_T$ is replaced
by the appropriate \textit{a posteriori} density $\pi_s(x)$. On
account of the relation $Z_T=X_T-\xi_s$ we have
\begin{eqnarray}
{\mathbb Q} [Z_T <z|\xi_s]={\mathbb Q} [X_T< z+\xi_s|\xi_s],
\end{eqnarray}
from which it follows that the conditional density of $Z_T$ is
given at time $s$ by $\pi_s(z+\xi_s)$. We can think of
$\pi_s(\xi_s+z)$ as a ``new'' \textit{a priori} density, now for the
random variable $Z_T$. Given this density we calculate the
conditional probability ${\mathbb Q}[Z_T <z|\eta_t]$ for
$t\in[s,T]$. By the method used to establish
Proposition~\ref{prop:4} and the probability law for the gamma
bridge $\{\delta_{tT}\}$ we deduce that the associated density
function is given by
\begin{eqnarray}
\frac{\rd}{\rd z}\,{\mathbb Q}[Z_T <z|\eta_t] ={\mathds
1}_{\{z>\eta_t\}} \frac{\pi_s(\xi_s+z) z^{1-m(T-s)}
(z-\eta_t)^{m(T-t)-1}} {\int_{\eta_t}^{\infty} \pi_s(\xi_s+z)
z^{1-m(T-s)} (z-\eta_t)^{m(T-t)-1}{\rm d} z}, \label{eq:8.80}
\end{eqnarray}
from which we see that the value process can be represented in the
following form:
\begin{eqnarray}
S_t = P_{tT}\left[\xi_s + \frac{\int_{\eta_t}^\infty
\pi_s(\xi_s+z) z^{2-m(T-s)}(z-y)^{m(T-t)-1} \rd
z}{\int_{\eta_t}^\infty \pi_s(\xi_s+z) z^{1-m(T-s)}
(z-y)^{m(T-t)-1} \rd z}\right] . \label{eq:ww01}
\end{eqnarray}
Making the substitution $z=x-\xi_s$ and also substituting
$\eta_t=\xi_t-\xi_s$, this expression reduces to the value process
obtained in (\ref{gammaAsset}).

\section{Example: gamma-distributed cash flow}
\label{sec:9}

When the terminal payout $X_T$ of the cumulative gains process
(\ref{eq:3.0}) is gamma distributed with mean $\kappa mT$ and
variance $\kappa^2mT$ for some choice of $\kappa$, the resulting
value process $\{S_t\}$ has an especially simple structure. In
particular, we are lead back to the ``${\mathbb Q}$-gamma'' model
discussed in the introduction. This can be seen as follows. Let
$\{\gamma_t\}$ be a standard gamma process with rate $m$, and let
$\{\gamma_{tT}\}$ be the associated gamma bridge. Then $X_T$ and
$\kappa \gamma_T$ have the same distribution; but since $\gamma_T$
and $\{\gamma_{tT}\}$ are independent, it follows that $\{X_T
\gamma_{tT}\}$ and $\{\kappa \gamma_T\gamma_{tT}\}$ have the same
probability law; therefore, $\{\xi_t\}$ has the same law as
$\{\kappa \gamma_t\}$, and hence is a ${\mathbb Q}$-gamma process,
with scale $\kappa$ and standard growth rate $m$. The fact that
$\xi_t$ is gamma distributed can be verified directly as follows.
The characteristic function of $X_T$ is $\phi_X(\lambda)=
(1-\ri\kappa\lambda)^{-mT}$. Substituting this into (\ref{eq:54-55})
and setting $z=(1-u)/ (1-\ri\kappa\lambda u)$, we deduce that
$\Phi_\xi(\lambda)= (1-\ri\kappa\lambda)^{-mt}$, which is the
characteristic function of a gamma distributed random variable with
mean $\kappa mt$ and variance $\kappa^2 mt$.

It is interesting to note that although the ${\mathbb Q}$-gamma
process has independent increments, the cumulative gains process
(\ref{eq:3.0}) has dependent increments. In particular, for the
covariance of $\xi_s$ and $\xi_t-\xi_s$ in the general case we have
\begin{eqnarray}
{\rm Cov}[\xi_s,\xi_t-\xi_s] = \frac{ms(t-s)}{T(mT+1)}\,{\mathbb E}
[X_T^2] - \frac{s(t-s)}{T^2}\left({\mathbb E} [X_T]\right)^2.
\end{eqnarray}
Hence a necessary condition for independent increments is given by
$({\mathbb E} [X_T])^2=mT\, {\rm Var}[X_T]$.

We conclude the paper by working out in some detail the value
processes for various claims in the ${\mathbb Q}$-gamma model. For
the density of $X_T$ we have $g_{\Gamma_T}(x)$, where
$g_{\Gamma_t}(x)$ is defined in (\ref{eq:ww2}). Substituting the
expression for the density function into (\ref{gammaAsset}) and
carrying out the relevant integration, we are led to the following
expression for the reserve process:
\begin{eqnarray}
S_t = P_{tT} \big( \xi_t + \kappa  m(T-t) \big). \label{eq:ww3}
\end{eqnarray}
Therefore, $\{S_t\}$ in this case is a linear function of
$\{\xi_t\}$. We observe that $S_0=P_{0T}\kappa mT$ and that
$S_T=X_T$, as required. An alternative derivation of
(\ref{eq:ww3}) is as follows. Since $\{\xi_t\}$ is a gamma process
with scale parameter $\kappa$ and standardised growth rate $m$, by
the Markov property we have $S_t=P_{tT}{\mathbb E}[ \xi_T|\xi_t]$,
and (\ref{eq:ww3}) follows immediately as a consequence of the
independent increments property of the gamma process.

These relations lead to simplifications in the valuation of
contingent claims. Let us work out, for example, the value $C_{tT}$
at time $t$ of a simple stop-loss reinsurance contract that pays out
$\max(X_T - K, 0)$ at $T$ for some fixed threshold $K$. In the
${\mathbb Q}$-gamma model we have
\begin{eqnarray}
C_{tT} = P_{tT}{\mathbb E} [(\xi_T -K)^+|\xi_t],
\end{eqnarray}
and hence by use of the independent increments property we deduce
that
\begin{eqnarray}
C_{tT} &=&\!P_{tT}\int_{(K-\xi_t)/\kappa}^\infty(\kappa z+ \xi_t-K)
\frac{z^{m(T-t)-1} \re^{-z}} {\Gamma[m(T-t)]}\, \rd z \nonumber \\
&=&\! P_{tT}\! \left[ \kappa \frac{\Gamma[m(T-t)+1, (K-\xi_t)
/\kappa]} {\Gamma[m(T-t)]} - (K-\xi_t) \frac{\Gamma[m(T-t),
(K-\xi_t)/\kappa ]} {\Gamma[m(T-t)]} \right],
\end{eqnarray}
where $\Gamma[a,z]=\int_z^\infty x^{a-1}\re^{-x}\rd x$ denotes the
incomplete gamma integral.

We proceed to calculate the associated Arrow-Debreu price $A_{st}$
in this model. By substituting (\ref{eq:ww3}) in (\ref{eq:zz5}) we
deduce that
\begin{eqnarray}
A_{st}(y) = P_{st} \frac{\kappa ^{-m(t-s)}}{\Gamma[m(t-s)]}\,
(y-\xi_s)^{m(t-s)-1} \exp\left(-\frac{1}{\kappa}(y-\xi_s)\right) .
\end{eqnarray}
It follows by use of (\ref{eq:ww3}) that the price at time $s$ of a
reinsurance contract with payout $(S_t-K)^+$ at $t$ is
\begin{eqnarray}
C_{st} &=& P_{st} {\mathbb E}_s[(S_t-K)^+] \nonumber \\ &=&
\int_0^\infty A_{st}(y) \left[ P_{t T}(y + \kappa m(T-t))-K\right]^+
\rd y \nonumber \\ &=& P_{sT}\left[ \frac{\Gamma[m(t-s)+1, \kappa^{-1}{
R}_s]} {\Gamma[m(t-s)]} - \kappa^{-1}R_s \frac{\Gamma[m(t-s),
\kappa^{-1}R_s]} {\Gamma[m(t-s)]} \right],
\end{eqnarray}
where $R_s=P_{t T}^{-1}K -(S_s+\kappa m(t-s))$.

\begin{acknowledgments}
The authors thank I.~Buckley, M.~Davis, E.~Hoyle, A.~Lokka,
D.~Madan, M.~Pistorius, and M.~Yor for stimulating discussions.
DCB acknowledges support from The Royal Society.  LPH and AM
acknowledge support from EPSRC grant number GR/S22998/01.
\end{acknowledgments}



\end{document}